%% file: arxiv.tex
\documentclass[manuscript, screen, noacm]{acmart}
\setcopyright{none}
\usepackage{csquotes}
\usepackage{bm}
\usepackage{makecell}
\usepackage{diagbox}
\usepackage{xspace}
\usepackage{siunitx}
\usepackage{placeins}
\usepackage{rotating}
\usepackage{adjustbox}
\usepackage{subcaption}
\usepackage{wrapfig}

\AtBeginDocument{%
  }

\newcommand{\matr}[1]{\bm{#1}}     

\newcommand{\vect}[1]{\bm{#1}}     

\newcommand{\ie}{\textit{i}.\textit{e}.\ }

\BeforeBeginEnvironment{wrapfigure}{\setlength{\intextsep}{0pt}}

\begin{document}

\title{Emergence of Nestedness in the Public Internet Peering Ecosystem}

\author{Justin Loye}
\authornote{All authors contributed equally to this research.}
\email{jloye@iij.ad.jp}
\affiliation{%
  \institution{Internet Initiative Japan Research Laboratory}
  \city{Tokyo}
  \country{Japan}
}
\orcid{0001-7797-6624}
\author{Sandrine Mouysset}
\authornotemark[1]
\email{sandrine.mouysset@irit.fr}
\affiliation{%
  \institution{Institut de Recherche en Informatique de Toulouse, Université Paul Sabatier}
  \city{Toulouse}
  \country{France}
}

\author{Katia Jaffrès-Runser}
\authornotemark[1]
\affiliation{%
  \institution{Institut de Recherche en Informatique de Toulouse, Toulouse INP-ENSEEIHT}
  \city{Toulouse}
  \country{France}}
\email{katia.jaffres-runser@irit.fr}

\renewcommand{\shortauthors}{Loye, Mouysset and Jaffrès-Runser.}

\begin{abstract}
    Nestedness is a property of bipartite complex networks that has been shown to characterize the peculiar structure of biological and economical networks. 
    In a nested network, a node of low degree has its neighborhood included in the neighborhood of nodes of higher degree.
    Emergence of nestedness is commonly due to two different schemes: i) mutualistic behavior of nodes, where nodes of each class have an advantage in associating with each other, such as plant pollination or seed dispersal networks; ii) geographic distribution of species, captured in a so-called biogeographic network where species represent one class and geographical areas the other one.
    Nestedness has useful applications on real-world networks such as node ranking and link prediction.

    Motivated by analogies with biological networks, we study the nestedness property of the public Internet peering ecosystem, an important part of the Internet where autonomous systems (ASes) exchange traffic at Internet eXchange Points (IXPs). 
    We propose two representations of this ecosystem using a bipartite graph derived from PeeringDB data. The first graph captures the AS [is member of] IXP relationship which is reminiscent of the mutualistic networks. The second graph groups IXPs into countries, and we define the AS [is present at] country relationship to mimic a biogeographic network.
    We statistically confirm the nestedness property of both graphs, which has never been observed before in Internet topology data. From this unique observation, we show that we can use node metrics to extract new key ASes and make efficient prediction of newly created links over a two-year period. 
    
\end{abstract}

\ccsdesc[500]{Computing methodologies~Model development and analysis}
\ccsdesc[500]{Networks~Public Internet}
\ccsdesc[500]{Networks~Topology analysis and generation}

\keywords{internet topology, nestedness, public peering}


\maketitle

\section{Introduction}

Nestedness is a graph property that characterizes the peculiar structure of certain real complex networks. These complex networks are composed of two classes of nodes, with interactions only observable between nodes of two different classes. They are thus modeled using a bipartite graph.
Such a bipartite complex graph is nested when, for any pair of nodes $(i,j)$ of the same class, if the degree of $i$ is greater than the degree of $j$, then the neighborhood of $j$ is included in the neighborhood of $i$.
This property has been observed mainly in two types of biological bipartite networks: $i)$ the so-called "mutualistic" networks \cite{bascompte2003nested}, such as pollination or seed dispersal networks, where the nodes of each class have a mutual benefit to associate with each other and $ii)$ the biogeographic networks representing the distribution of species over geographical sites \cite{patterson1986nested}.
Only more recently has this structural property been identified in non-biological networks: in the international trade network \cite{bustos2012dynamics}, where it accounts for the ability of countries to produce products of increasing complexity, and in online communication systems \cite{borge2017emergence} as well.

Here, by making an analogy with mutualistic and biogeographic networks, we question for the first time to our knowledge the concept of nestedness in data describing the topology of the Internet. The nestedness property being the result of a dynamic process of evolution, it can be leveraged the identify key players and predict the evolution of the network structure to some extent.

At the coarsest granularity, the Internet is composed of autonomous systems (ASes) that can be, for example, content providers, Internet service providers, or network service providers.
They essentially interconnect with each other in a client-to-provider or peer-to-peer relationship, thanks to infrastructures that can be private or public.
A significant amount of Internet traffic between ASes occurs in public infrastructures known as Internet Exchange Points (IXPs) \cite{chatzis2013benefits}. IXPs are generally non-profit organizations that operate in metropolitan areas with the objective of facilitating the peering interconnections of the ASes that are members, in exchange for a fee from the ASes to cover their operating costs. Together, ASes and IXPs form what is known as the \enquote{public peering ecosystem}.

There is no single or complete view of the Internet topology, making it difficult to understand the Internet at the AS level.
The Internet is notoriously difficult to measure \cite{motamedi2014survey, roughan201110}, and information is not disclosed by its constituents for competitiveness and security reasons.
Current AS-level topology data \cite{giotsas2014inferring} is known to miss most peering links \cite{ager2012anatomy, giotsas2014inferring} due to their invisible nature \cite{oliveira2008search}, even though these links are a growing part of the Internet thanks to developments in the public peering ecosystem in recent years \cite{bottger2018elusive}.
The dataset generally used by the scientific community to study the evolution of public peering is PeeringDB \cite{peeringdb}.
PeeringDB contains self-reported data about ASes, IXPs, and the membership of ASes at IXPs.
This database is expressive enough to identify the major players in the peering ecosystem \cite{bottger2018looking, loye2022global}, and can be modeled as a bipartite graph representing the AS [is a member of] IXP relationship \cite{nomikos2017re}.
PeeringDB is being collected on a daily basis by CAIDA \cite{lodhi2014}, allowing for temporal studies.

In this work, we study the nestedness of the public Internet peering ecosystem by defining two graphs from the PeeringDB dataset.
First, we know that ASes and IXPs gain mutual benefit from their association, with the former ones gaining access to novel ASes and the latter ones gaining attractiveness.
Thus, we construct the AS [is member of] IXP graph to capture the essence of a mutualistic biological network.
Second, we construct another graph on the basis of an analogy with the biogeographic networks.
Suppose an AS expands its business to a new geographic area.
Then, it is more likely that the AS will establish itself in a certain order over time: starting with the most densely populated or strategically important places and then expanding to less populated or strategic places.
This results in a process of selective immigration similar to the one causing nestedness in biogeographic networks (cf. \autoref{fig:nestedness}).
We mimic such graph by grouping IXPs into countries, and defining an AS [is present at] country graph.
If the analogies are relevant, we expect to find nested networks that, through the applications of the nestedness concept such as node ranking and link prediction, improve our understanding of the public peering ecosystem.

Our study makes the following contributions:
\begin{itemize}
    \item We reveal the nested structure of the topology of two networks representing the public peering ecosystem using PeeringDB;
    \item We identify new key peering actors from metrics related to nestedness. These actors are for instance high degree ASes able to access hard to reach areas of the world;
    \item Following a 27-month temporal study, we show that nestedness can be leveraged to effectively predict new link creation.
\end{itemize}
Depending on the acceptance of the paper, we provide the code to reproduce the results presented hereafter and we make the tools implemented easy to use for the community.

This paper is organized as follows. In \autoref{sec:background}, we present background on the public peering ecosystem and nestedness. In \autoref{sec:modeling}, we extract two graphs from PeeringDB and detail the necessary preconditioning for a nestedness study. The nestedness of both graphs is then assessed in \autoref{sec:assessment}. Finally, we study in \autoref{sec:usecases} two applications of nestedness: new key ASes identification and link prediction.

\section{Background}
\label{sec:background}
\subsection{Mathematical notations}
Next, we consider that a graph $G(E,V)$ is composed of a set of edges or links $E$, and a set of vertices or nodes $V$.
It is described by an adjacency matrix $\matr{A}$.
A graph is said to be bipartite if all links are shared between two disjoint subsets of nodes, and if the graph is not directed, it can be fully described by its bi-adjacency matrix $\matr{B}$.
We denote by $\vect{d}(i)$ the degree of a node $i$, i.e. its number of links.
A graph is said to be connected if there exists a succession of links connecting every pair of nodes.
\subsection{PeeringDB}
PeeringDB is the authoritative source of information about the public peering ecosytem.
It is a non-profit and user-maintained database made to facilitate interconnections of ASes at IXPs.
It consists in self-reported information about ASes (ASN, business type, traffic levels, ratio of inbound or outbound traffic,\ldots), IXPs (facilities, country,\ldots), and the membership of ASes at IXPs.
As of March 1st, 2021, 869 IXPs and 21388 ASes are registered.

Although it is a self-reported database, PeeringDB is considered to be up-to-date and correct.
The data is now mostly uploaded automatically and it is in ASes best interests to give correct information.
As stated in \cite{bottger2018looking}, \enquote{[PeeringDB] has a very good standing in the network operators community, which naturally has very big interest in having reliable peering information available}.
The authors give examples of important organisations that rely on PeeringDB like Google, Netflix and Cloudflare.  
Comparing three years of PeeringDB snapshot with BGP data, authors in \cite{lodhi2014} show that \enquote{PeeringDB membership is reasonably representative of the Internet’s transit, content, and access providers in terms of business types and geography of participants}.
Thus, in line with these previous works, we consider PeeringDB as ground truth.

However, it's important to note that PeeringDB only tells us about the public peering ecosystem, which is only a sub-part of the inter-domain Internet. Although IXPs are an important part of today's Internet, enabling local traffic exchange, tier 1 ASes bypassing \cite{bottger2018elusive} and CDN deployment \cite{openconnect}, PeeringDB does not report on private network interconnects (PNIs). PNIs are a preferred means of interconnection for ASes with the means to do so. For example, Tier-1s use PNIs because, for them, public peering means losing a customer. This is also the case for CDNs: Facebook (Meta) favors connections via PNI rather than IXP for reasons of monitoring and congestion control \cite{schlinker2019internet}.

\subsection{Nestedness}
\begin{figure}[tbh]
    \centering
    \includegraphics[width=0.96\textwidth]{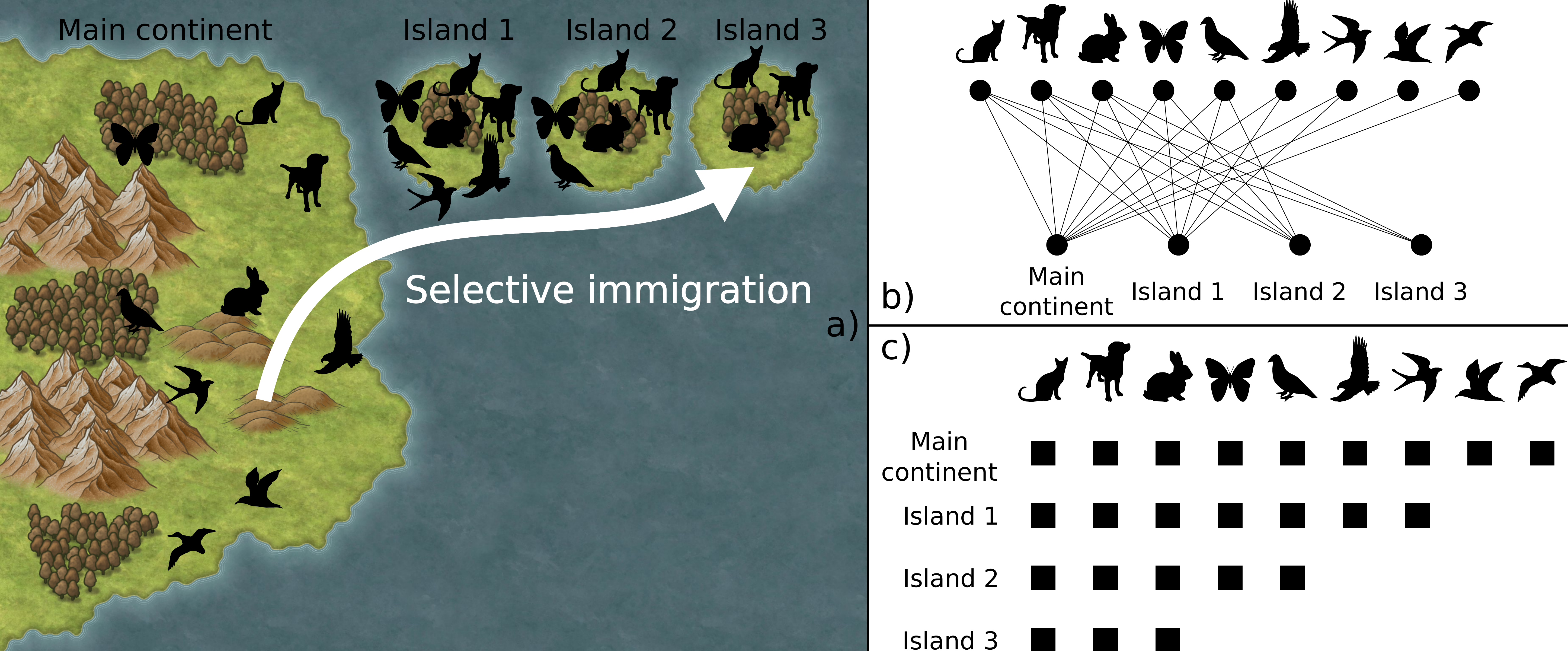}
    \caption{Illustration of a perfectly nested biogeographic network. a) The selective immigration process causing the nested structure b) Bipartite biogeographic network c) A possible permutation of the rows and columns of the bi-adjacency matrix (here ordered by decreasing degree) showing the characteristic triangular structure.}
    \vspace{-4mm}
    \label{fig:nestedness}
\end{figure}
Nestedness is a graph property commonly found in real world data such as biological and economical networks.
Formally, a graph is perfectly nested when, for any pair of nodes $(i,j)$, if the degree of $i$ is greater than the degree of $j$, then the neighborhood of $j$ is included in the neighborhood of $i$.
If the graph is bipartite, as is often the case in nestedness studies, then the same condition holds for any pair of nodes within the same class (cf. \autoref{fig:nestedness}-(b)).
As a consequence, it is possible to find an arrangement of rows and columns such as its bi-adjacency matrix presents a triangular shape on the top left corner (cf. \autoref{fig:nestedness}-(c)). Such a structure is representative of a perfectly nested network. For a recent nestedness literature overview, we refer the reader to \cite{mariani2019nestedness}.

Nestedness is a property characterized at different levels.
At the graph-level, it consists in quantifying how much the graph is close to a perfectly nested graph.
After that, one might be interested in node-level nestedness that consists in finding the arrangement of rows and columns allowing to reveal the triangular structure in the adjacency matrix.
This ordering represents a new ranking of the nodes in terms of their individual contribution to nestedness.
And finally, one might look for nestedness at the partition-level to identify communities where nodes arrange themselves in a nested manner.

Nestedness was first conceptualized to explain the geographic distribution of animal species resulting from a selective immigration process.
Such an immigration process is illustrated in \autoref{fig:nestedness}-(a): a set of species living on a main continent immigrates to islands by crossing only small distances of water.
As a result, the species farthest from the main continent are necessarily a subset of the nearest species, inducing the nested structure.
This structure has been first observed in a biogeographic network representing the distribution of mammalians in archipelagos \cite{patterson1986nested}.
This distribution is best represented by an unweighted bipartite graph, with nodes in classes \enquote{mammalians} or \enquote{archipelagos} and a link signifying the presence of species at geographical sites.
It was then observed in other biological bipartite networks where the nodes of each class mutually benefit from their association.
These so-called "mutualistic" networks are for example plant pollination networks or seed dispersal networks. In these two examples, the nodes of the first class are plants whose reproduction is facilitated, while the nodes of the second class are animals finding food.

Since then, nestedness has drawn the attention of network scientists for its useful applications.
In a nested network, the node ordering w.r.t. to their individual contribution to nestedness is the best strategy of attack to disconnect the network \cite{dominguez2015ranking}. Moreover, new links are more likely to appear in the characteristic upper triangular structure, therefore it can be leveraged for link prediction \cite{bustos2012dynamics}.

\section{Modeling PeeringDB for a nestedness study}
\label{sec:modeling}
Motivated by analogies with biological nested networks, we study the nestedness of two networks derived from PeeringDB data. 
Since ASes and IXPs derive mutual benefit from their association, the AS [is member of] IXP graph is reminiscent of a mutualistic biological network.
The other analogy stems from an intuition on the way ASes expand their reach in the network to new geographic areas. They may start with densely populated areas or strategically important places to maximize revenue, and later develop to less populated or less strategic places. This results in a process of selective immigration similar to the biogeographic network illustrated in \autoref{fig:nestedness}). We mimic such graphs by grouping IXPs into countries, and defining an AS [is present at] country graph.

In order to analyze the nestedness property of both graphs, it is necessary to ensure that the graphs are well conditioned, i.e. that their adjacency matrix respects the same properties as the ones manipulated in previous works on nestedness.

\subsection{Constraints on the bi-adjacency matrix}

The nested biological networks commonly studied in the literature are present in the database \textit{web of life} \cite{weboflife}. Most of them are bipartite, with bi-adjacency matrix presenting the following properties \cite{bruno2020ambiguity}:
\begin{itemize}
    \item The ratio between the number of rows $N_r$ and columns $N_c$ is balanced (there is an equilibrium between the number of nodes in both classes).
    As a consequence, the eccentricity $e = \left| N_r - N_c \right|/(N_r + N_c)$ should be between $0$ and $0.5$. \label{enum:equilibrium}
    \item The matrix has a small size $N = N_r + N_c$ (at most 1000 nodes in both classes).
    As a consequence, numerous tools related to the evaluation of nestedness do not scale up. \label{enum:size}
    \item The matrix is neither too sparse nor too dense.
    Since nestedness is based on the comparison of overlapping neighborhoods, graphs with many low-degree nodes are penalized when detecting nestedness, even if these low-degree nodes do not play an important role in the overall structure of the graph. The scarce studies involving large networks therefore filter out the nodes with the fewest connections \cite{palazzi2019online, palazzi2021ecological}.
    We stress that filtering nodes is also a widespread practice in other fields. For example the bipartite user [has rated] movie graph from the movielens100k dataset \cite{harper2015movielens}, widely used to test recommendation algorithms, contains only users with a degree superior to 20.
\end{itemize}
In the rest of this section, we present how the AS [is member of] IXP and AS [is present at] country graphs are built and conditioned to meet these requirements.

\subsection{AS-IXP graph}
\begin{figure}
    \centering
    \includegraphics[width=0.48\textwidth]{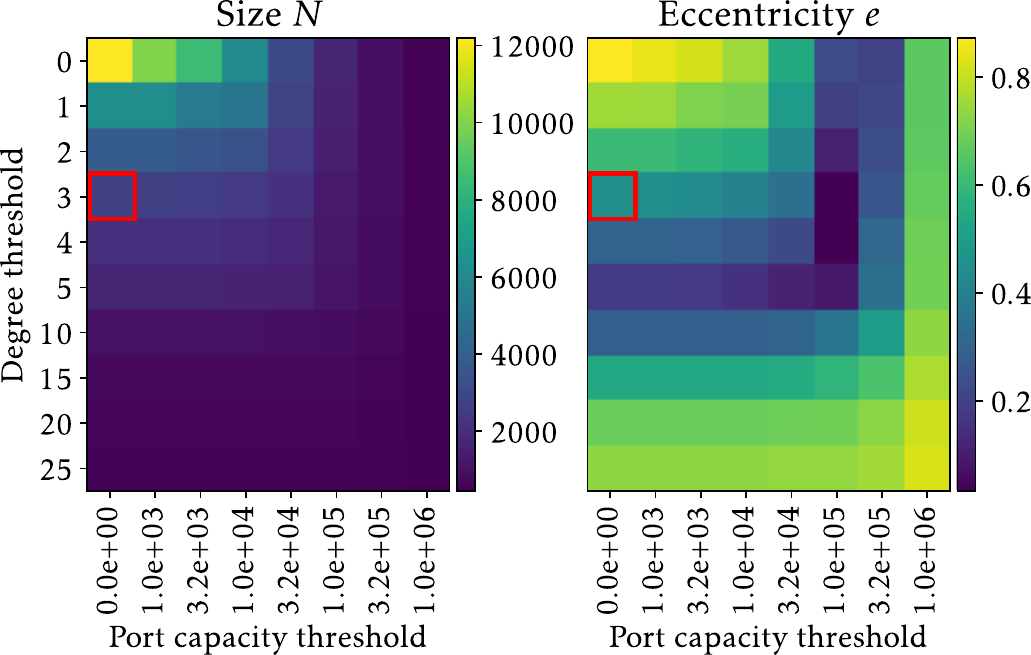}
    \caption{Impact of filtering on properties of the AS-IXP graph. The chosen filter $d \leq 3$ is highlighted in red.}
    \vspace{-6mm}
    \label{fig:filtering}
\end{figure}
The AS [is member of] IXP graph, referred to as {\it AS-IXP graph} in the remainder of this paper, is a bipartite graph composed of 11407 ASes reporting 31889 associations to 780 IXPs.
It is obtained with PeeringDB membership data as proposed in \cite{nomikos2017re}, for which we weight the links with the port size attribute.
We consider only the main connected component, which comprises more than 99\% of the total nodes.

To meet the constraints outlined earlier, we study in \autoref{fig:filtering} the evolution of adjacency matrix properties (eccentricity and network size $N$) according to node filters based on degree and weighted degree (we use for the later the terminology \enquote{port capacity} introduced in \cite{bottger2018looking} as the sum of the port sizes of an AS' links).
As highlighted in red in \autoref{fig:filtering}, we conclude to filter out the nodes with a degree $d \leq 3$.
The main reason for this is to reduce the eccentricity to 0.5 while having the least restrictive filter possible.
The chosen filter has the advantage as well of making the network small enough for a nestedness study, with 2681 nodes.
It reduces the number of nodes with low degree and therefore penalizes the nestedness detection, which is particularly necessary since half of the ASes in PeeringDB have a single connection.
We do not apply a filter on the port capacity because its effect is less pronounced than the one of the filter on degrees, and could result in the removal of high degree ASes with low associated port sizes. 
The resulting graph presents 1942 ASes connected to 739 IXPs with 17689 links.

\subsection{AS-country graph}
The AS [is present at] country is referred to as the {\it AS-country graph} in the following. 
We construct the AS-country graph by merging, in the unfiltered AS-IXP graph, the IXPs by their reported countries.
We obtain the presence of 11407 ASes in 112 countries, inducing a strong imbalance between both classes.
To have a balanced ratio, we select the most central ASes in the graph according to three metrics. First, we select the top 100 ASes in decreasing order of degree in the AS-IXP graph. Second, we select the top 100 ASes in decreasing order of PageRank and reverse PageRank in a weighted and directed graph model of the AS-IXP graph as proposed in \cite{loye2022global}. In this paper, authors leverage the router's port size and the reported traffic balance of ASes (Heavy or Mostly Inbound or Outbound, Balanced) of PeeringDB to build a weighted and directed bipartite graph that captures the communication capacity of each link. PageRank (resp. reverse PageRank) metric applied to this graph can be used to identify nodes concentrating (resp. emitting) traffic.
The final set of ASes is the union of the three top 100 sets. The resulting set is composed of 182 ASes.

In the following analysis, we only work with the unweighted version of the two AS-country and AS-IXP graphs. We do this for reasons of convenience: i) nestedness studies are mainly carried out on unweighted graphs, and are therefore more mature ii) many nestedness tools used don't take weights into account iii) to simplify the scope of our analysis (for example, in the link prediction study, predicting the link weight is more difficult than predicting the presence of a new link). From an application point of view, this restriction amounts to working only on the absence and presence of ASes, which we argue is still valuable information for stakeholders.

\section{Nestedness assessment}
\label{sec:assessment}
In this section, we characterize for the first time to our knowledge the presence of nestedness in Internet topology data.
We first introduce the methods used to assess nestedness that are based on shared best practices outlined in \cite{ulrich2009consumer},\cite{beckett2014falcon} and \cite{mariani2019nestedness}.

\subsection{Detecting nestedness}

\paragraph{Statistical significance of metrics}

To evaluate the nestedness of a network, it is considered good practice to use several metrics. Most metrics do not include a null model and return a value that, in absolute terms, has no clear interpretation.
It is therefore necessary to compare this value to a null ensemble consisting of a large number of nestedness measures on random networks sharing properties with the network to be evaluated.
We generate the null ensembles with the null model Proportional-Proportional (PP) that is commonly used since it offers a balance between errors of type I (identifying non-nested networks as nested) and type II (rejecting nested networks) \cite{mariani2019nestedness}.
This model is also used in community detection in the popular modularity measures of Newman \cite{girvan2002community} and its bipartite counterpart \cite{barber2007modularity}.
State-of-the-art implementations of this model are known as \enquote{Bascompte's PP} \cite{bascompte2003nested} and \enquote{corrected PP} \cite{palazzi2021comprehensive}, the latter being a correction of the widely used Bascompte's PP aiming to further reduce the number of type I errors.
Random networks generated by those models have nodes that share on average the same degree as the nodes of the network to test.

Once a null ensemble is generated, the nestedness statistical significance of the network to be evaluated is assessed with the $p$-value and the $z$-score \cite{ulrich2009consumer}.
The $p$-value is the fraction of the null ensemble with higher nestedness than the test network.
If there is no network in the null ensemble that is more nested than the network to be tested, then we conservatively set the value of $p$ to be the inverse of the ensemble size, as proposed in \cite{beckett2014falcon}.
A $p$-value less than 0.05 is considered sufficient to say that the network is more nested than the statistical set \cite{beckett2014falcon}. On the other hand, $p>0.95$ indicates an anti-nested pattern.
The $z$-score is the difference between the test network measure and the mean of the null ensemble measures, normalized by its standard deviation.

\paragraph{Graph-level metrics}
Graph-level nestedness metrics quantify how much a network deviates from a perfectly nested structure.
They leverage the structure of the $N_r \times N_c$ bi-adjacency matrix, where $N_r$ and $N_c$ are respectively the number of nodes in both classes.
By convention, we will refer to nodes in one class as the \enquote{row nodes}, the others being the \enquote{column nodes}.
In the following, we report the three main metrics whose descriptions are given in the reference \cite{mariani2019nestedness}.

The first metric we consider is the popular metric called {\it Nestedness metric based on Overlap and Decreasing Fill}, noted NODF \cite{NODF}.
Given a pair of row nodes $(i,j)$ with degrees such that $\vect{d}(i)>\vect{d}(j)$, we expect that
their number of common neighbors $\matr{O}_{i,j} = \sum_\alpha \matr{A}_{i,\alpha} \matr{A}_{j,\alpha}$ is equal to $\vect{d}(j)$ for a perfectly nested network, or smaller than $\vect{d}(j)$ otherwise.
The total NODF value $\eta$ is calculated from the NODF value of each class.
The NODF value per row $\eta^R$ is defined by:
\begin{equation}
    \eta^R = \sum_{i,j}\frac{\matr{O}_{i,j}}{\vect{d}(j)}\Theta\left(\vect{d}(i)-\vect{d}(j)\right),
\end{equation}
where $\Theta$ is the Heaviside function, defined by $\Theta(x) = 1$ if $x>0$, $\Theta(x)=0$ if $x \leq 0$.
This function allows to count only the pairs of $(i,j)$ row nodes such that $\vect{d}(i)>\vect{d}(j)$.
The maximum contribution of a pair is 1. Thus, the maximum value of $\eta^R$ is equal to the number of pairs of row nodes $N_r(N_r-1)/2$, with $N_r$ the number of row nodes.
In the same way, the NODF metric per column $\eta^C$ is given by : 
\begin{equation}
    \eta^C = \sum_{\alpha,\beta}\frac{\matr{O}_{\alpha,\beta}}{\vect{d}(\beta)}\Theta\left(\vect{d}(\alpha)-\vect{d}(\beta)\right),
\end{equation}
where $\matr{O}_{\alpha,\beta} = \sum_i \matr{A}_{i,\alpha} \matr{A}_{i,\beta}$. The maximum attainable value is $N_c(N_c-1)/2$, with $N_c$ the number of columns.
Finally, the nestedness value of a network is given by the total NODF which is the ratio between the NODF value computed on the graph $\eta^R + \eta^C$ and the maximum attainable value:
\begin{equation}
    \eta = \frac{\eta^R + \eta^C}{ \frac{N_r(N_r-1)}{2} + \frac{N_c(N_c-1)}{2}}.
    \label{ch1:eq:NODF}
\end{equation}\par

The second metric we consider is a variant of NODF taking into account a null model proposed in \cite{sole2018revealing} :


\begin{equation}
    \tilde{\eta} = \frac{2}{N_r+N_c} \biggl\lbrace \sum_{i,j}\frac{\matr{O}_{i,j}-\langle \matr{O}_{i,j}\rangle}{(N_r-1)\vect{d}(j)} \Theta(\vect{d}(i)-\vect{d}(j))
    +  \sum_{\alpha, \beta} \frac{\matr{O}_{\alpha,\beta}-\langle \matr{O}_{\alpha,\beta}\rangle }{(N_c-1)\vect{d}(\beta)} \Theta(\vect{d}(\alpha)-\vect{d}(\beta)) \biggr\rbrace,
    \label{ch1:eq:eta}
\end{equation}

with the expectation of the number of common neighbors $\langle\matr{O}_{i,j}\rangle$ given by 
considering that the neighbors of each node are drawn randomly, respecting their degree.
In this case, the probability that the two nodes have a common link is
$\langle \matr{O}_{i,j}\rangle = \vect{d}(i)\vect{d}(j)/N_c^2$.
The same reasoning holds for any pair of $\alpha, \beta$ nodes of the other class.
If the network is not more nested than the null model, then $\tilde{\eta} \leq 0$ ,
and the closer $\tilde{\eta}$ is to its maximum value of 1, the more nested the network is.\par

Lastly, we consider the metric known as spectral radius.
 It has been shown in \cite{bell2008graphs} that the spectral radius $\rho$ of a matrix, i.e. the eigenvalue with the highest modulus,
is maximal when this matrix is perfectly nested.
Staniczenko et al. \cite{staniczenko2013ghost} statistically study the behavior of $\rho$ on matrices with imperfect nestedness,
to conclude that $\rho$ can be used to quantify the nestedness of real complex networks.

\subsection{Nestedness in AS-country graph}

The results of the statistical tests of nestedness for the AS-country graph are given in \autoref{ch3:tab:PeeringDima_graph_level_measurements}.
The $p$-value, associated with a high $z$-score, shows that no null ensemble provided a more nested network than the AS-IXP graph. We can conclude that the AS-country network is highly nested.

Although this is outside the scope of our work, we have also evaluated the nestedness of the weighted version of this graph, which takes into account the sum of port sizes that an AS has in a country.
Our goal is to assess that the nested structure is not an artefact caused by making the graph unweighted, i.\ e.\ by the binarization of the adjacency matrix. 
It is indeed possible for a binary bi-adjacency matrix to be nested, but to have a weight distribution on the triangular structure following a non-nested or even an anti-nested pattern (for an illustration we refer the reader to Fig.\ 2 in \cite{staniczenko2013ghost}).
To do this, we use the methodology introduced in \cite{staniczenko2013ghost}, which consists in choosing a metric and a null ensemble suitable for weighted graphs.
We therefore use the spectral radius metric on the weighted adjacency matrix, and also measure it on an ensemble of 10000 random graphs generated by a null model keeping the source and destination of the original links fixed but randomly permuting the weights between links.
We find that the graph is strongly nested, as no random graph is more nested than it (p-value of $0.0001$) and the z-score is high at $11.16$.
We can therefore conclude that the observed nestedness is not an artefact caused by making the graph unweighted.

\subsection{Nestedness in AS-IXP graph}
\label{subsec:assess_nest_ixp}
For the AS-IXP graph, the results presented in \autoref{ch3:tab:netcut_graph_level_measurements} are quite mixed.
Among the six combinations of null ensembles and metrics, four indicate a strongly nested network with $p<0.05$ while two identify the network as strongly anti-nested with $p$ close to $0.95$.
Thus, we cannot conclude on the nested structure of the AS-IXP graph.

\begin{table}[htbp]
    \caption{Statistical test results, presented as the $p$ value and $z$ score in parentheses, of the graph-level nestedness.} 
    \vspace{-3mm}
    \begin{minipage}[t]{0.5\linewidth} 
        \centering
        \begin{subtable}{\linewidth}
            \centering
            \caption[Statistical test results for AS-country graph]{AS-Country}
            \input{tables/PeeringDima_graph_level_measurements.tex}
            \label{ch3:tab:PeeringDima_graph_level_measurements}
        \end{subtable}
    \end{minipage}%
    \begin{minipage}[t]{0.5\linewidth} 
        \centering
        \begin{subtable}{\linewidth}
            \centering
            \caption[Results of statistical tests for AS-IXP graph]{AS-IXP}
            \input{tables/netcut_graph_level_measurements.tex}
            \label{ch3:tab:netcut_graph_level_measurements}
        \end{subtable}
    \end{minipage} 
    \label{tab:comparison}
\end{table}

However, it has been shown in \cite{loye2022global} that the IXPs membership network exhibits a community structure that correlates with the geographic proximity of ASes and IXPs.
Thus, one might wonder if nestedness should not rather be sought in the network's communities, i.e.\ at the level of the partition. To do so, we leverage a metric recently developed in \cite{sole2018revealing} to quantify the presence of nested communities, also called \enquote{blocks} due to the resulting structure in the adjacency matrix. 

\subsection{In-Block Nestedness}
This metric called In-Block Nestedness (IBN) is an adaptation of the metric $\tilde{\eta}$ (\autoref{ch1:eq:eta}) also introduced in \cite{sole2018revealing}. It can be optimized in the same way as the classical modularity \cite{girvan2002community, fortunato2010community} to obtain a network partition.

We first define the number of nodes in the same class and community of a row node $i$
by $C_i=\sum_j\delta(g(i), g(j))$, where $g(i)$ refers to the community of a node $i$ and $\delta$ is the Kronecker delta, i.e.\ $\delta(i,j)=1$ if $i=j$, $\delta=0$ otherwise.
The same equation holds for a column node $\alpha$: $C_\alpha=\sum_\beta\delta(g(\alpha), g(\beta))$.
The number of common neighbors now includes community membership:
\begin{equation}
    \matr{O}_{i,j} = \sum_{\alpha =1}^{N_r} \matr{A}_{i,\alpha}\matr{A}_{j,\alpha}\delta(g(i), g(\alpha)), \qquad \matr{O}_{\alpha,\beta} = \sum_{i=1}^{N_c} \matr{A}_{i,\alpha}\matr{A}_{i,\beta}\delta(g(i), g(\beta)).
\end{equation}
The value of IBN $\tilde{I}$ is then
\begin{multline}
    \tilde{I} = \frac{2}{N_r+N_c}\left\{ \sum_{i,j}^{N_r} \left[\frac{\matr{O}_{i,j}-\langle \matr{O}_{i,j}\rangle}{\vect{d}(j)(C_i-1)} \Theta(\vect{d}(i)-\vect{d}(j))\delta(g(i),g(j)) \right] \right.\\
    +\left.  \sum_{\alpha,\beta}^{N_c}\left[\frac{\matr{O}_{\alpha, \beta}- \langle \matr{O}_{\alpha,\beta}\rangle}{\vect{d}(\beta)(C_\alpha-1)} \Theta(\vect{d}(\alpha)-\vect{d}(\beta))\delta(g(\alpha),g(\beta))\right]\right\}.
    \label{ch1:eq:ibn}
\end{multline}

In order to find nested communities, the authors of \cite{sole2018revealing} optimize $\tilde{I}$ with the extremal optimization (EO) method \cite{boettcher2002optimization, duch2005community}.
In the present work, we use the implementation of this algorithm made available by the authors of \cite{palazzi2019macro} in \cite{cosin3_uoc_2021_4557009}.
We will compare this nested community partition to the classical notion of strongly connected community partition, the latter being obtained by optimizing Barber's modularity \cite{barber2007modularity} with the same implementation of the EO algorithm \cite{palazzi2019macro,cosin3_uoc_2021_4557009}.

The composition of the largest communities obtained by modularity and IBN optimization is given in \autoref{ch3:tab:bloc_composition_mod} and \ref{ch3:tab:bloc_composition_ibn} respectively.
We can observe that the modularity tends to partition the network in communities of similar sizes.
The 15 hypergiant ASes identified in \cite{bottger2018looking}\footnote{namely Apple Inc, Amazon.com, Facebook, Google Inc.\, Akamai Technologies, Yahoo!, Netflix, Hurricane Electric, OVH, Limelight Networks Global, Microsoft, Twitter Inc., Twitch, Cloudflare and Verizon Digital Media Services.}, \ie ASes leveraging IXPs to have a global reach, are split between communities 0 ($40\%$) and 7 (60\%), which does not capture their actual global footprint on the Internet.
For the IBN on the contrary, a large portion of the nodes and all of the hypergiants are grouped in a  single community. If we consider the port capacity metric, introduced in \cite{bottger2018looking} as the sum of all port sizes reported in PeeringDB for each AS, this community is even more predominant, containing ASes that account for 77\% of total port capacity. The connected subgraph induced by this community is highly nested, with the results presented in \autoref{ch3:tab:netcut_graph_level_measurements_bloc1} showing the lowest possible $p$-value and a high $z$-score. Motivated by these observations, we will study in the following this community which we call the main nested component of the AS-IXP graph.
\begin{table*}[]
    \caption[Community composition of the modularity optimization by the extremal optimization algorithm.]{Community composition of the modularity optimization by the extremal optimization algorithm.}
    \vspace{-2mm}
    \resizebox{0.8\textwidth}{!}{%
\begin{tabular}{@{}ccccccc@{}}
    \toprule
    Community &
    \begin{tabular}{c} ASes\\proportion (\%)\end{tabular} &
    \begin{tabular}{c} IXPs\\proportion (\%)\end{tabular} &
    \begin{tabular}{c} Port capacity\\ proportion (\%)\end{tabular} &
    \begin{tabular}{c} Unique\\countries\end{tabular} &
    Countries distribution \\ \midrule

    6  & 19.0 & 8.9  & 11.2 & 21 & FR: 14 | GB: 9 | IT: 8 | CZ: 5 | BG: 4   \\
    0  & 15.1 & 16.0 & 36.4 & 25 & US: 68 | HK: 5 | TZ: 5 | CA: 4 | PH: 4   \\
    10 & 11.7 & 7.3  & 4.0  & 3  & ID: 25 | DE: 15 | UA: 14                 \\
    4  & 11.3 & 11.0 & 1.9  & 27 & AU: 15 | DE: 12 | CN: 5 | NL: 5 | RO: 5  \\
    1  & 11.0 & 11.9 & 9.3 & 12 & BR: 36 | US: 29 | CA: 8 | EC: 3 | AO: 2  \\
    7  & 9.6  & 21.0 & 25.4 & 71 & IN: 18 | PL: 11 | JP: 10 | US: 9 | ES: 6 \\
    16 & 7.0  & 7.7  & 4.3  & 9  & RU: 34 | US: 7 | LT: 4 | EE: 3 | LV: 3   \\
    12 & 6.5  & 5.7  & 2.5  & 9  & AU: 24 | NZ: 6 | CL: 3 | BO: 2 | CN: 2   \\ \bottomrule
\end{tabular}}
\label{ch3:tab:bloc_composition_mod}
\vspace{-2mm}
\end{table*}
\begin{table*}[]
    \caption[Same as \autoref{ch3:tab:bloc_composition_mod} for communities obtained by the optimizing In-Block Nestedness $\tilde{I}$.]{Same as \autoref{ch3:tab:bloc_composition_mod} for communities obtained by the optimization of In-Block Nestedness $\tilde{I}$.}
    \vspace{-2mm}
    \resizebox{0.8\textwidth}{!}{%
    \begin{tabular}{@{}ccccccc@{}}
    \toprule
    Community &
    \begin{tabular}{c} ASes\\proportion (\%)\end{tabular} &
    \begin{tabular}{c} IXPs\\proportion (\%)\end{tabular} &
    \begin{tabular}{c} Port capacity\\ proportion (\%)\end{tabular} &
    \begin{tabular}{c} Unique\\countries\end{tabular} &
    Countries distribution \\ \midrule
        1 & 44.4 & 45.6 & 77.3 & 100 & US: 86 | DE: 14 | CA: 12 | IN: 11 | GB: 9  \\
        98  & 7.0 & 3.1  & 3.0  & 15 & ID: 3 | US: 3 | BO: 2 | DE: 2 | GB: 2      \\
        31  & 5.6 & 1.9  & 2.3  & 3  & AU: 12 | AR: 1 | US: 1                  \\
        38  & 5.3 & 4.6  & 4.0 & 2  & BR: 33 | PT: 1                          \\
        19  & 5.1 & 3.0  & 1.1  & 3  & ID: 20 | AR: 1 | HK: 1                  \\
        14  & 5.1 & 4.1  & 2.3  & 4  & RU: 26 | UA: 2 | PH: 1 | UZ: 1          \\
        53  & 5.0 & 2.2  & 0.3  & 11 & DE: 4 | CH: 2 | US: 2 | CA: 1 | CN: 1   \\ \bottomrule
        \end{tabular}}
    \label{ch3:tab:bloc_composition_ibn}
    \vspace{-2mm}
\end{table*}
\begin{table}
    \caption[Same as \autoref{tab:comparison} for the community 0 of the AS-IXP graph.]{
    Same as \autoref{tab:comparison} for the community 0 of the AS-IXP graph.}
    \vspace{-2mm}
    \centering
    \begin{tabular}{@{}llll@{}}
        \toprule
        \theadfont\diagbox[width=9em]{Null\\model}{Nestedness\\metric}
                   & $\tilde{\eta}$      & NODF          & spectral radius \\ \midrule
        PP (Bascompte) & 0.001 (144.36)  & 0.001 (113.21)  & 0.001 (56.31)   \\
        PP (corrected) & 0.001 (27.97)   & 0.001 (19.53)   & 0.001 (16.79)   \\
    \end{tabular}
    \vspace{-3mm}
    \label{ch3:tab:netcut_graph_level_measurements_bloc1}
\end{table}

\section{Use cases}
\label{sec:usecases}
Nestedness has been exhibited for the AS-country and AS-IXP networks, but the triangular structure in the bi-adjacency matrix which is characteristic of nestedness is yet to be revealed. Such a structure is obtained by a relevant ordering of the rows and columns of the bi-adjacency matrix, and has several applications:
\begin{itemize}
    \item The derivation of a new and original ranking of nodes that offers a novel interpretation of their importance in the network;
    \item The prediction of link appearance in the network based on the idea that a new link is more likely to appear in the triangular structure of the matrix.
\end{itemize}

In this section, we investigate both applications for the two PeeringDB networks.
Thus, we first present the node-level nestedness metrics derived by unveiling the triangular graph structure following recent literature methods \cite{mariani2019nestedness}.
It allows us to identify new important nodes of the peering ecosystem.
Next, we show how leveraging this triangular structure allows the effective prediction of newly created links.

\subsection{Node-level metrics}
The simplest node-level metric we can leverage to re-order rows and columns of the bi-adjacency matrix to reveal nestedness is the node degree.
It is a classical metric fulfilling the decreasing degree condition of nestedness, and has a simple interpretation. 
However, its definition does not take into account the second and only other  property of nestedness: the neighborhoods' overlap.

Another possible ordering of nodes is given by the popular state-of-the-art BINMATNEST algorithm \cite{rodriguez2006new}. The authors propose a graph-level nestedness metric sensitive to the nodes ordering in the matrix. A genetic algorithm is deployed to maximize this graph-level metric by swapping the rows and columns. The ranking of nodes is given by the row or the column indices observed in the optimal isomorphism. This metric clearly aims at accounting for the overlap between neighborhoods property. 

Lastly, we introduce the fitness-complexity metric that also produces state-of-the art ordering \cite{lin2018nestedness}, but which has the advantage over BINMATNEST of having a clear interpretation and a simple and efficient implementation.
It is in the context of economic complexity, first introduced by economists to analyze the country [export] product bipartite network, that the fitness-complexity metric came into being \cite{tacchella2012new}.
This approach views the export of a product as the result of a production process that requires a country to have all the capabilities necessary to manufacture the product. In this view, countries with more capacity are more competitive in world trade, as they have the ability to produce and export more products than countries with less capacity.
In addition, a notion of complexity is added to products to show that they can only be produced by competitive countries.
This notion is itself taken into account for competitiveness: a competitive country exports complex products.
This approach is modeled by coupled, non-linear and recursive fitness and complexity equations.
The fitness equation measures the economic health of a country based on the quality of its exported products and the complexity equation measures the quality of the products based on the economic health of the countries that export them and the number of countries that export them.
For a network described by the adjacency matrix $\matr{A}$, the {\bf fitness} $\vect{f}(i)$ of a country $i$ is defined as the
sum of the complexity scores $\vect{q}(\alpha)$ of the products $\alpha$ exported by this country:
\begin{equation}
    \vect{f}(i) = \sum_\alpha \matr{A}_{i,\alpha}\vect{q}(\alpha), \quad \vect{q}(\alpha) = \frac{1}{\sum_i \matr{A}_{i,\alpha}/\vect{f}(i)}.
\end{equation}
The scores of the products depend on the number of countries that export them but also on the economic health (fitness) of the countries that produce them.
A competitive country generally exports complex products, a less developed country exports products that are easier to produce.
The complexity measure is based on these assumptions.
To solve these recursive and coupled sequences, we compute iteratively :
\begin{equation}
\begin{split}
    \vect{f}_n(i) &= \sum_\alpha \matr{A}_{i,\alpha}\vect{q}_{n-1}(\alpha), \quad
    \vect{q}_n(\alpha) = \frac{1}{\sum_i \matr{A}_{i,\alpha}/\vect{f}_{n-1}(i)}, \label{eq:FQ_scores}
\end{split}
\end{equation}
beginning with uniform non-null vectors $\vect{f}_0$ and $\vect{q}_0$.
At each iteration, these vectors are divided by their mean.
For a better convergence, we use a slightly modified version of the complexity in \autoref{eq:FQ_scores} as proposed in \cite{stojkoski2016impact}:
\begin{equation}
    \vect{q}_n(\alpha) = \frac{1}{\sum_i \matr{A}_{i,\alpha}(1-\vect{f}_{n-1}(i))}. \label{eq:Q_scores_m}
\end{equation}
Ordering countries and products by these rankings reveals the triangular structure inherent to nestedness.
Since its publication, this algorithm has been used outside the context of economics, notably in biology, to measure the nestedness of nodes and to rank them from most to least nested.

\subsection{New key players}

We have seen that there are several nestedness metrics at the node level. In addition to the traditional degree metric which satisfies the first nestedness condition, namely a degree decrease, there are two specific nestedness metrics known as BINMATNEST and fitness-complexity. After verifying that all metrics reveal the characteristic triangular structure of nestedness, we ask here whether BINMATNEST and fitness-complexity provide an original view compared to the degree by identifying new key ASes, and what this teaches us about the public peering ecosystem.

\begin{figure}
    \centering
    \includegraphics[width=0.7\textwidth]{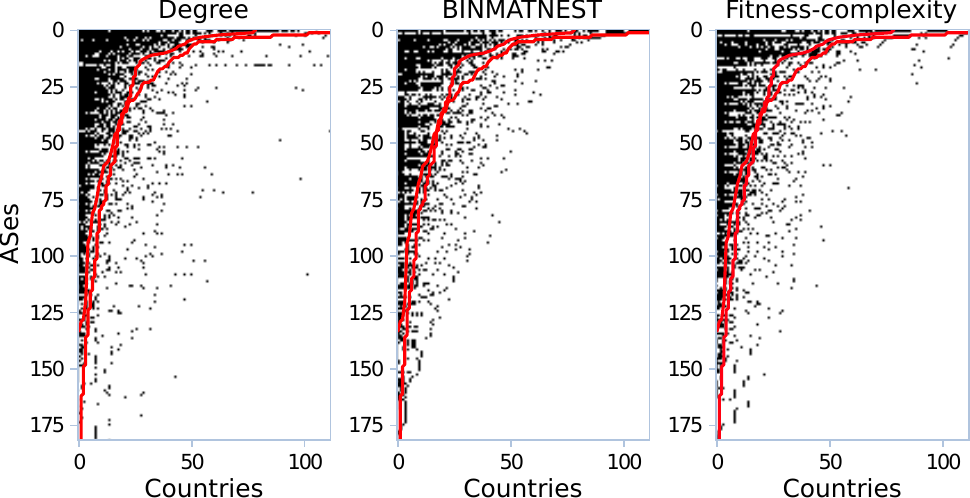}
    \caption[Ordering of the bi-adjacency matrix of the AS-country graph for different node-level nestedness metrics.]{
        Ordering of the bi-adjacency matrix of the AS-country graph for different node-level nestedness metrics.}
    \vspace{-4mm}
    \label{ch3:fig:PeeringDima_adjacency_packed}
\end{figure}

We show the AS-country bi-adjacency matrix ordered by different node-level nestedness metrics in \autoref{ch3:fig:PeeringDima_adjacency_packed}.
All metrics reveal a triangular structure characteristic of nestedness. The links are concentrated in a stylized upper left triangle, with a convex hypothenuse caused by the filling rate of the matrix. The fill rate is the ratio between the number of existing links and the number of possible links, i.e. $\phi = \sum_{i,j}\matr{B}_{i,j}/(N_r\times N_c)$. Depending on its value, the hypothenuse of a perfectly nested biadjacency matrix is either a straight line ($\phi = 0.5$ ), convex ($0\leq\phi<0.5$) or concave ($0.5<\phi\leq 1$). To visualize how marked the AS-IXP triangular structure is, we compare it to a perfectly nested network with the same number of links. The hypothenuse of a perfectly nested network is given by the diversity-ubiquity curves \cite{bustos2012dynamics}. The diversity and ubiquity curves are respectively the curves giving the degree of the AS and the countries. They are represented in red on the figure. We can see that the density of the links is indeed higher on the left of these curves than on the right, thus giving a visual indication of nestedness for all of the metrics considered.

We find a difference between the degree metric on one side, and the BINMATNEST and fitness-complexity metrics on the other side.
Degree ordering has a dense triangle of links but some of them are far to the right of the diagonal, while the other two have no links to the right of the diagonal at the cost of a less dense triangle.
This means that the fitness-complexity and BINMATNEST rankings, specific to nestedness, provide a new ranking that is original compared to the traditional degree one.
The disagreements in the rankings are reported in \autoref{ch3:tab:nestedness_rankings} that lists the top 30 ASes for all three rankings.
New players characterized by a large gap between both their nestedness ranking and their degree ranking appear.
Facebook's secondary ASN (now Meta), that is used to distribute its most popular content \cite{facebookpeering} efficiently through offnet deployment as close as possible to
end-users \cite{gigis2021seven},
ranks significantly better according to BINMATNEST and fitness, with +13 and +14 increase respectively over the degree.
It enters the top 4 AS, and ahead of Facebook's main AS, which may seem peculiar.
Other AS have a significant increase in ranking, such as network service providers ROSTELECOM (+80 and +91) and NETIX communication (+16 and +32).
According to the fitness-complexity metric, these new players are interpreted as being able to access hard-to-reach areas of the world, which can not be captured in degree ranking.

Although hard to validate, we believe most of these results are consistent from an operational point of view. Facebook's secondary AS is used to distribute its most popular content in cached form, which we assume is essential in remote areas where bandwidth is limited. NETIX communication offers global peering solution and advertise its \enquote{global reach} \cite{netix}, so they might be specialized in hard-to-reach regions. In the case of ROSTELECOM, we attribute the large increase to a low degree coupled with its presence in both Europe and Asia.

We also report the appearance of new actors in the AS-IXP graph with \autoref{ch3:tab:NestComp_fitness_betweenness_rankings}\footnote{We do not show results from BINMATNEST due to performance issues.}.
Among them, we find operators related to domain resolution (DNS) such as RIPE NCC K-Root Operations (improvements over the degree of +30), DNS-OARC-112 (+29), Netnod (+11).
The RIPE NCC K-Root Operations and Netnod ASes are associated with the 13 root name servers and thus play a structuring role in the global Internet, which is clearly revealed in this new ranking. These AS are well ranked because they extend their presence to isolated IXPs.

\begin{figure}
    \centering
    \includegraphics[width=0.4\textwidth]{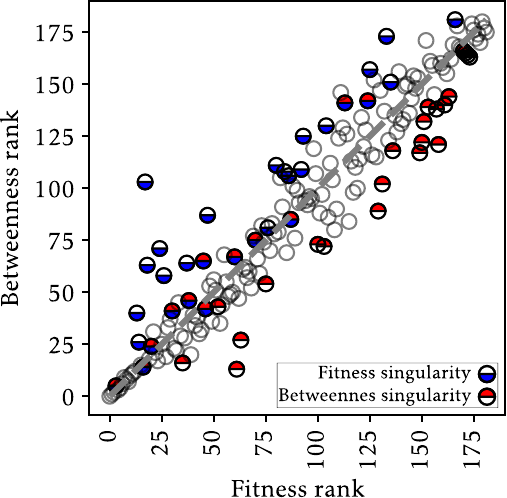}
    \caption{Correlation between the fitness and betweenness rankings of nodes in the AS-country graph. Singularities, defined as a ranking deviation of at least 10 with respect to the degree, are highlighted.}
    \vspace{-4mm}
    \label{fig:singularities}
\end{figure}
To assert the originality of the fitness metric, we compare it with the betweenness metric proposed in \cite{freeman1977set}, which is fundamental in network science while being original relative to the degree metric.
The betweenness centrality quantifies the proportion of shortest paths between all pairs of nodes in a network that pass through a particular node, indicating its importance as a bridge between sub-structures of the network \cite{holme2002attack}. We plot against each other in \autoref{fig:singularities} the AS ranks based on the betweenness and the fitness metrics.
First of all, we see that the two rankings are correlated, with points distributed along the diagonal.
However, when we highlight the ASes with a deviation of at least 10 between one of their rankings and the ranking by degree, which we call singularity, we notice that these singularities are generally located on either side of the diagonal.
So, while some AS are singular according to both metrics, the majority are singular according to only the betweenness or the fitness.
This shows that fitness is original to both degree and betweenness.

We point out that all the AS rankings presented differ greatly from a CAIDA reference ranking based on ASes customer cone \cite{luckie2013relationships}. This is mainly due to the fact that ASes with many customers, such as backbone ASes with no provider known as "tier 1 networks", are less interested in public peering. For them, settlement-free peering means losing a potential customer.
Thus, we see the limits and specificity of the PeeringDB dataset. Now that we have shown that nestedness metrics reveal the triangular structure of the PeeringDB networks and give an original view of the public peering ecosystem, we will investigate if the presence of a triangular structure enables efficient link prediction.

\begin{table}[htbp]
\caption{Comparison of node-level nestedness rankings to the traditional degree ranking. The ranking difference with respect to degree is shown in square brackets.} 
\vspace{-2mm}
    \begin{minipage}[t]{0.5\linewidth} 
        \centering
        \begin{subtable}{\linewidth}
            \centering
            \caption[Nestedness rankings for AS-country graph]{AS-Country}
\input{tables/PeeringDima_nodes_top30_tab.tex}
            \label{ch3:tab:nestedness_rankings}
        \end{subtable}
    \end{minipage}%
    \begin{minipage}[t]{0.5\linewidth} 
        \centering
        \begin{subtable}{\linewidth}
            \centering
            \caption[Nestedness rankings for AS-IXP graph]{AS-IXP}
            \resizebox{0.98\textwidth}{!}{%
            \begin{tabular}{@{}llll@{}}
                \toprule
                   & Degree                    & Fitness                                                       \\ \midrule
                1  & Hurricane Electric        & Hurricane Electric {[}0{]}                     \\
                2  & Cloudflare                & Packet Clearing House {[}1{]}                \\
                3  & Packet Clearing House     & Packet Clearing House AS42 {[}1{]}            \\
                4 &
                  Packet Clearing House AS42 &
                  Cloudflare {[}-2{]} \\
                5  & Google LLC                & Akamai Technologies {[}2{]}               \\
                6  & Microsoft                 & Google LLC {[}-1{]}                       \\
                7  & Akamai Technologies       & Microsoft {[}-1{]}                        \\
                8  & Facebook Inc              & Facebook Inc {[}0{]}                      \\
                9  & Amazon.com                & Amazon.com {[}0{]}                        \\
                10 &
                  Netflix &
                  VeriSign Global Registry Services {[}5{]}  \\
                11 &
                  Yahoo! &
                  EdgeCast Networks (VDMS) {[}1{]} \\
                12 &
                  EdgeCast Networks (VDMS) &
                  Netflix {[}-2{]}  \\
                13 & Limelight Networks Global & Limelight Networks Global {[}0{]}                              \\
                14 & Fastly, Inc.              & Fastly, Inc. {[}0{]}                                      \\
                15 &
                  VeriSign Global Registry Services &
                  Yahoo! {[}-4{]}  \\
                16 & G-Core Labs               & \textbf{Netnod} {[}11{]}                                 \\
                17 & Apple Inc.                & Equinix Internet Exchange - MLPE {[}25{]}                  \\
                18 & Riot Games                & i3D.net {[}1{]}                                      \\
                19 & i3D.net                   & \textbf{RIPE NCC K-Root Operations} {[}30{]}         \\
                20 & Blizzard Entertainment    & Salesforce.com {[}14{]}                             \\
                21 & Subspace                  & Netskope {[}2{]}                                  \\
                22 & SG.GS                     & Subspace {[}-1{]}                            \\
                23 & Netskope                  & \textbf{DNS-OARC-112} {[}29{]}                       \\
                24 & Imperva                   & Riot Games {[}-6{]}                          \\
                25 & Twitch                    & Apple Inc. {[}-8{]}                          \\
                26 & Valve Corporation         & G-Core Labs {[}-10{]}                     \\
                27 & Netnod                    & Twitch {[}-2{]}                            \\
                28 & Swisscom                  & Imperva {[}-4{]}                           \\
                29 &
                  SoftLayer Technologies, Inc. &
                  Quantil Networks {[}8{]}  \\
                30 & StackPath (Highwinds)     & Blizzard Entertainment {[}-10{]}        \\ \bottomrule
            \end{tabular}}
            \label{ch3:tab:NestComp_fitness_betweenness_rankings}
        \end{subtable}
    \end{minipage} 
    \vspace{-4mm}
    \label{tab:nestedness_comparison}
\end{table}

\subsection{Link prediction}
As a more concrete application, we ask whether the nestedness observed in PeeringDB graphs can be used to predict the appearance or disappearance of new links, which is motivated by the fact that nestedness captures the attractiveness of IXPs and countries for ASes looking for novel interconnection opportunities. To assess the quality of the proposed link prediction scheme, the temporal evolution of the network has to be extracted as a ground truth. Thus, we first explain how temporal snapshots of the two graphs are extracted from PeeringDB. Second, we present the prediction method, and describe how the predictions are validated for both graphs. Results are given for both graphs numerically, together with an illustration for the AS-country graph.

\paragraph{Networks preparation}
The temporal study is made possible by CAIDA \cite{lodhi2014}, which provides the community with daily PeeringDB snapshots from 2013 to present time.
Our study window is the period between 2019-01-01 and 2021-03-01, and we choose a time resolution of a month, leading to 27 snapshots of the two graphs.\\
Since the graphs evolve over time, we study here the largest connected component present over the time window of interest. Thus, we select the nodes of both graphs by considering only the sub-graphs induced by the nodes present in all snapshots. This selection leads to only a few nodes being disconnected from the main connected component, that are removed as well.
For the AS-country graph, this results in a network of 181 ASes and 91 countries.
For the AS-IXP graph of which we previously studied only the main nested component, we perform a new largest nested community search as described in \autoref{subsec:assess_nest_ixp}, this time at the beginning of the time period.
To be comprehensive, our goal is to start from a community and study its evolution, without the knowledge of the future community structure. Once again, we find a main community strongly nested that is similar to the one presented in \autoref{ch3:tab:bloc_composition_ibn}: it includes 51.8\% of ASes, 39.7\% of IXPs, 80\% of all port capacity, 55 countries and all 15 hypergiants. We therefore consider only the sub-graph induced by this community.\\
For the AS-country and AS-IXP graphs, respectively, the number of links varies from 1863 to 2064 and 6329 to 6893.
The links persist over time: for both graphs, a mean of 98\% of links of a snapshot are still present in a snapshot taken 6 months later.

\paragraph{Link prediction models}
We adopt the method of link prediction proposed in \cite{bustos2012dynamics}.
This method is used to predict both the appearance and disappearance of links in nested networks.
We only present how to predict appearances for the sake of simplicity, the extension to the prediction of disappearances being straightforward and given in \cite{bustos2012dynamics}.
It consists in i) ordering all unformed links at the beginning of the time period according to their likelihood of appearance ii) predicting the links sequentially according to this order. 

The first point is achieved by taking advantage of the triangular structure of the bi-adjacency matrix.
Let us consider a probit model of parameters $\alpha, \beta, \gamma$ that is fit from the nested arrangement of links at the beginning of the time period:
\begin{equation}
    \matr{A}_{i,j} = \alpha \vect{d}(i) + \beta \vect{d}(j) + \gamma(\vect{d}(i) \times \vect{d}(j)) + \matr{\epsilon}_{i,j}
    \label{eq:probit}
\end{equation}
The probit residual term $\matr{\epsilon}$ represents the deviation from the nested arrangement of links and the model.
A low negative value corresponds to an unformed link that was expected to be present in the nested structure. It is therefore a link that is likely to be created.
On the other hand, high positive value corresponds to a formed link not expected to be present, and thus being likely to disappear in later snapshots.

\paragraph{Validation}
We assess the quality of the predictor by doing the following checks. Based on the first snapshot and its matrix, we define the order of link appearance (resp. disappearance) following the decreasing (resp. increasing) order of the probit model of \eqref{eq:probit}. For each prediction, we check in the subsequent snapshots if the link has appeared (resp. disappeared) or not.

The quality of the predictor is then assessed by the receiver operating characteristic (ROC) curve \cite{bradley1997use}. The true positive rate (TPR, predicted links actually created or deleted afterwards) and the false positive rate (FPR, predicted links not created or not deleted afterwards) is evaluated at each prediction.
The evolution of the TPR as a function of the FPR gives the ROC curve, and the area under this curve (AUC) contained in the unit square quantifies the quality of the predictor: 1 for the best predictor, 0.5 for a random predictor, and 0 for the worst predictor.

\begin{figure}[b]
    \centering
    \includegraphics[width=0.8\textwidth]{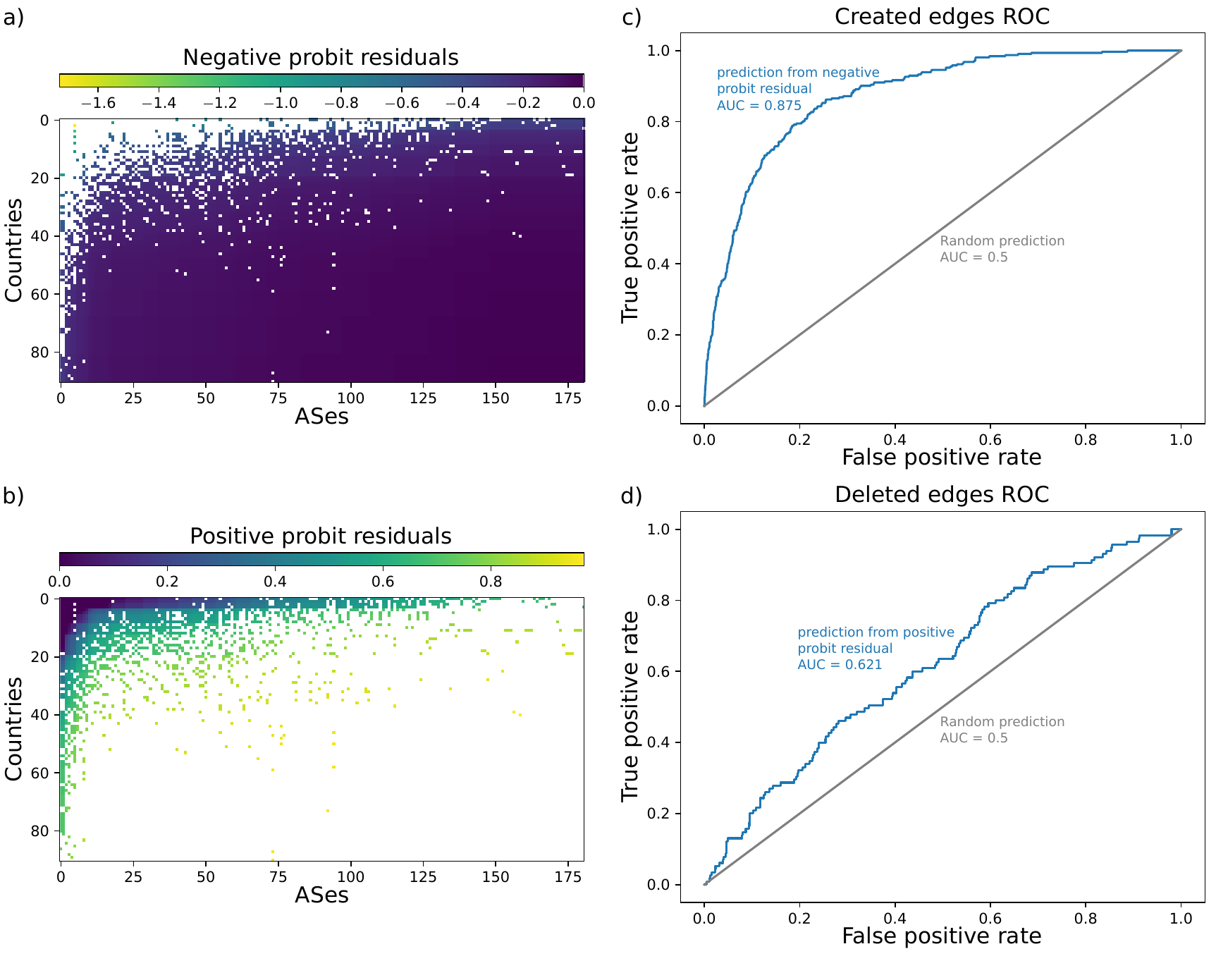}
    \caption{Link prediction for the AS-country graph.
    a) Negative probit residuals, i.e. the likelihood for links to appear (links already present and therefore excluded of the prediction are marked in white).
    b) Positive probit residuals, i.e. the likelihood for links to disappear (links already missing and therefore excluded of the prediction are marked in white). 
    c) ROC of the prediction of newly created links.
    d) ROC of the prediction of newly deleted links.}
    \label{fig:peeringdima_link_pred}
\end{figure}

\begin{figure}[b]
    \centering
    \includegraphics[width=0.8\textwidth]{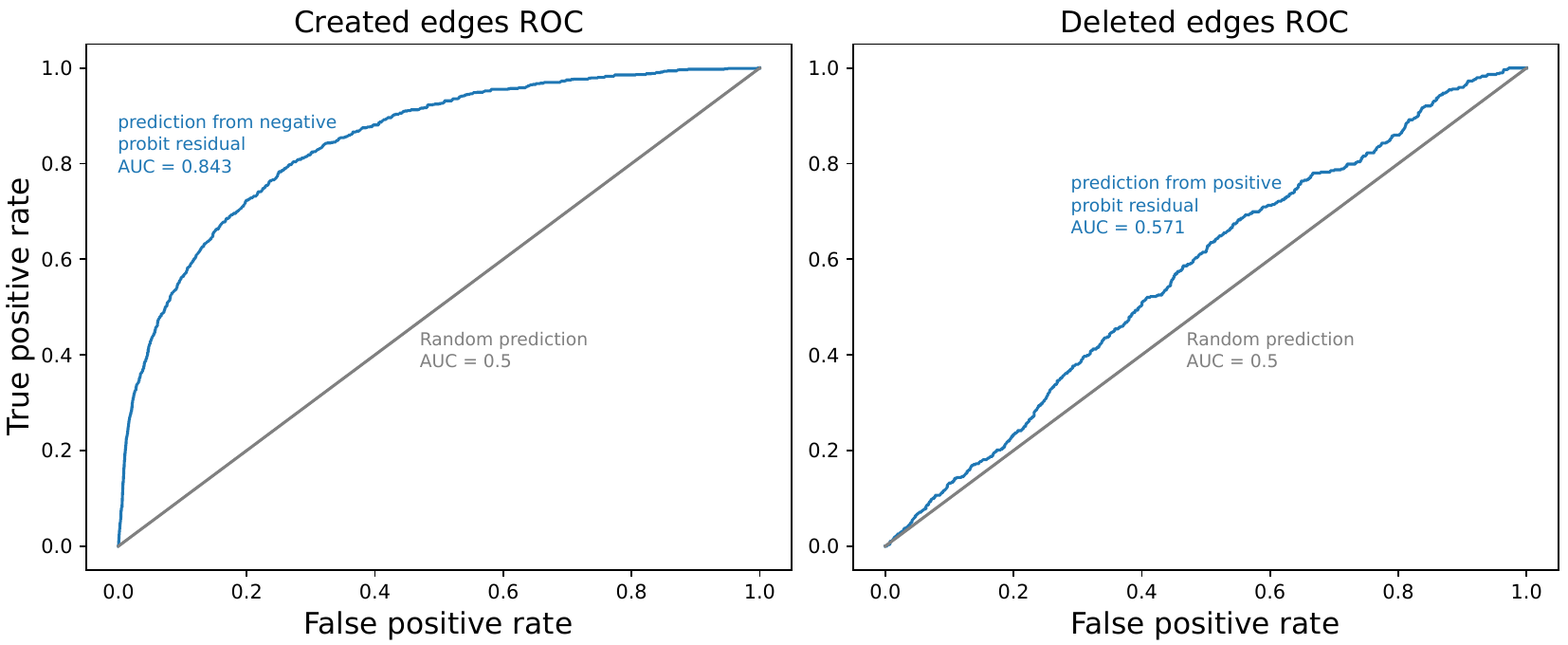}
    \caption{Link prediction for the AS-IXP graph.
    Left panel: ROC of the prediction of newly created links.
    Right panel: ROC of the prediction of newly deleted links.}
    \vspace{-4mm}
    \label{fig:netcut_link_pred}
\end{figure}

\paragraph{Results}
The results of the link prediction for the AS-country graphs are shown in \autoref{fig:peeringdima_link_pred}. We see in a) that links with negative probit residuals and therefore likely to appear are located inside the triangular structure. These links are proposed first by the predictor, leading to a high predictor quality with an AUC of the ROC curve shown in b) of 0.875, far outperforming a random prediction. The positive probit residuals shown in b) are located outside of the triangle. This time, the prediction based on the nested structure is not significantly better than a random one, with an AUC of the ROC curve shown in d) of 0.621.
The same predictor behavior, good for created links and average for deleted links, is found for the AS-IXP graph.
The two ROC curves for creation and deletion are shown in \autoref{fig:netcut_link_pred}.
The first has an AUC of 0.88 and the second 0.55.
Overall, these results show that the nestedness observed in the PeeringDB ecosystem can be efficiently leveraged to predict the creation of new links.
This could serve as an IXP recommender system for network operators.

\section{Conclusion}
In this work, we have drawn a parallel with biological ecosystems to reveal the nestedness of graph representations of the public peering ecosystem. To the best of our knowledge, this is the first time that nestedness has been observed in Internet topology data. We have shown that nestedness has concrete applications to better understand the public peering ecosystem, an ecosystem that is difficult to grasp because of the lack of topology data due to its confidential nature and measurement limitations. The node-level nestedness metric known as fitness-complexity provides a new interpretable ranking of ASes.
This metric highlights the ability of ASes to reach isolated regions of the world and IXPs, which is not captured with the traditional degree and betweenness metrics.
With a 27-month time study, we also show that nestedness effectively predicts new associations between ASes and countries in the world, as well as new membership of ASes to IXPs.\par
We believe these applications will be useful to stakeholders in the public peering ecosystem.
For example, link prediction can be seen as an IXP recommender system for network operators.
Future work will explore the use of other analysis tools based on deep learning such as graph neural networks to better understand the structure of the Internet.

\begin{acks}
We would like to thank Dima Shepelyansky, Marc Bruyère and anonymous reviewers for their helpful comments during the development of this work.
\end{acks}
 \FloatBarrier
\newpage
\bibliographystyle{ACM-Reference-Format}
\bibliography{ref.bib}



\end{document}

%% file: tables/PeeringDima_graph_level_measurements.tex
\resizebox{0.98\textwidth}{!}{%
\begin{tabular}{@{}lccc@{}}
    \toprule
\theadfont\diagbox[width=9em]{Null\\model}{Nestedness\\metric}&
                          $\tilde{\eta}$      & NODF           & spectral radius \\ \midrule
    PP (Bascompte) & 0.001 (80.59)  & 0.001 (61.04)  & 0.001 (29.97)   \\
    PP (corrected) & 0.001 (27.34)  & 0.001 (20.93)  & 0.001 (11.42)   \\ \bottomrule
\end{tabular}} 

%% file: tables/netcut_graph_level_measurements.tex
\resizebox{0.98\textwidth}{!}{%
    \begin{tabular}{@{}llll@{}}
    \toprule
    \theadfont\diagbox[width=9em]{Null\\model}{Nestedness\\metric}
                   & $\tilde{\eta}$      & NODF          & spectral radius \\ \midrule
    PP (Bascompte) & 0.001 (103.94) & 0.001 (86.61) & 0.001 (94.58)   \\
    PP (corrected) & 0.945 (-1.63)  & 0.977 (-1.94) & 0.001 (39.68)   \\ \bottomrule

    \end{tabular}}

%% file: tables/PeeringDima_nodes_top30_tab.tex
\resizebox{0.98\textwidth}{!}{%
\begin{tabular}{@{}llll@{}}
\toprule
 &
  Degree &
  Binmatnest &
  Fitness  \\ \midrule
1 &
  Packet Clearing House &
  Packet Clearing House {[}0{]} &
  Packet Clearing House {[}0{]} \\
2 &
  Packet Clearing House AS42 &
  Packet Clearing House AS42 {[}0{]} &
  Packet Clearing House AS42 {[}0{]} \\
3 &
  Cloudflare &
  \textbf{Facebook Inc AS63293} {[}14{]} &
  Cloudflare {[}0{]} \\
4 &
  Akamai Technologies &
  Cloudflare {[}-1{]} &
  \textbf{Facebook Inc AS63293} {[}13{]} \\
5 &
  Hurricane Electric &
  Akamai Technologies {[}-1{]} &
  Akamai Technologies {[}-1{]} \\
6 &
  VeriSign Glob. Reg. Serv. &
  VeriSign Glob. Reg. Serv. {[}0{]} &
  Hurricane Electric {[}-1{]} \\
7 &
  Microsoft &
  Hurricane Electric {[}-2{]} &
  VeriSign Glob. Reg. Serv. {[}-1{]} \\
8 &
  Facebook Inc &
  Google LLC {[}1{]} &
  RIPE NCC K-Root Ops {[}5{]} \\
9 &
  Google LLC &
  Netnod {[}2{]} &
  Netnod {[}2{]} \\
10 &
  Amazon.com &
  Facebook Inc {[}-2{]} &
  Facebook Inc {[}-2{]} \\
11 &
  Netnod &
  RIPE NCC K-Root Ops {[}2{]} &
  Google LLC {[}-2{]} \\
12 &
  Subspace &
  Microsoft {[}-5{]} &
  Microsoft {[}-5{]} \\
13 &
  RIPE NCC K-Root Ops &
  Amazon.com {[}-3{]} &
  Amazon.com {[}-3{]} \\
14 &
  Imperva &
  Subspace {[}-2{]} &
  \textbf{NetIX Com. Ltd.} {[}32{]} \\
15 &
  Netflix &
  Netskope {[}12{]} &
  Netskope {[}12{]} \\
16 &
  Anexia &
  Imperva {[}-2{]} &
  Subspace {[}-4{]} \\
17 &
  Facebook Inc AS63293 &
  G-Core Labs {[}2{]} &
  Quantil Networks {[}11{]} \\
18 &
  Riot Games &
  Quantil Networks {[}10{]} &
  \textbf{Rostelecom} {[}91{]} \\
19 &
  G-Core Labs &
  Telstra (International) {[}34{]} &
  NORDUnet {[}54{]} \\
20 &
  Yahoo! &
  Swisscom {[}5{]} &
  Anexia {[}-4{]} \\
21 &
  Valve Corporation &
  Anexia {[}-5{]} &
  CacheFly {[}29{]} \\
22 &
  SG.GS &
  Riot Games {[}-4{]} &
  Swisscom {[}3{]} \\
23 &
  Zenlayer Inc &
  Netflix {[}-8{]} &
  Riot Games {[}-5{]} \\
24 &
  Limelight Networks Global &
  Wnet Telecom USA {[}41{]} &
  Valve Corporation {[}-3{]} \\
25 &
  Swisscom &
  Zenlayer Inc {[}-2{]} &
  Twitter, Inc. {[}42{]} \\
26 &
  M247 &
  Valve Corporation {[}-5{]} &
  Netflix {[}-11{]} \\
27 &
  Netskope &
  Yahoo! {[}-7{]} &
  Telstra (International) {[}26{]} \\
28 &
  Quantil Networks &
  M247 {[}-2{]} &
  Imperva {[}-14{]} \\
29 &
  EdgeCast Networks (VDMS) &
  \textbf{Rostelecom} {[}80{]} &
  Yahoo! {[}-9{]} \\
30 &
  Equinix IX - MLPE &
  \textbf{NetIX Com. Ltd.} {[}16{]} &
  M247 {[}-4{]} \\ \bottomrule
\end{tabular}}

